\documentclass[aps, pre, twocolumn, final, floatfix, showpacs, letterpaper, superscriptaddress]{revtex4}

\usepackage[utf8]{inputenc}
\usepackage[english]{babel}
\usepackage[T1]{fontenc}

\usepackage{graphicx}
\usepackage{amsmath}

\renewcommand{\phi}{\varphi}
\renewcommand{\rho}{\varrho}
\renewcommand{\epsilon}{\varepsilon}
\renewcommand{\theta}{\vartheta}
\newcommand{\dd}{\mathrm{d}}
\newcommand{\mrX}{\mathrm{X}}
\newcommand{\mrY}{\mathrm{Y}}
\newcommand{\bbx}{\mathbf{x}}
\newcommand{\bby}{\mathbf{y}}
\newcommand{\Th}{\text{Th}}
\newcommand{\Cx}{\text{Cx}}
\newcommand{\mS}{\mathcal{S}}

\begin{document}

\title{Inferring   coupling strength from event-related 
dynamics}

\author{Szymon Łęski} 
\affiliation{Department of Neurophysiology, Nencki Institute of
  Experimental Biology, ul. Pasteura 3, 02-093 Warszawa, Poland}
\author{Daniel K. Wójcik}
\affiliation{Department of Neurophysiology, Nencki Institute of
  Experimental Biology, ul. Pasteura 3, 02-093 Warszawa, Poland}
\affiliation{Warsaw School of Social Psychology, ul. Chodakowska
  19/31, 03-815 Warszawa, Poland}
\email{s.leski@nencki.gov.pl} 
\email{d.wojcik@nencki.gov.pl} 
\date{\today}
\begin{abstract} 
  We propose an approach for inferring strength of coupling between
  two systems from their transient dynamics.  This is of vital
  importance in cases where most information is carried by the
  transients, for instance in evoked potentials measured commonly in
  electrophysiology.  We show 
  viability of our approach using nonlinear and linear measures of
  synchronization on a population model of thalamocortical loop and on
  a system of two coupled Rössler-type oscillators in non-chaotic regime.
\end{abstract}
\maketitle

\section{Introduction}

Coherent actions of apparently distinct physical systems often provoke
questions of their possible interactions.  Such coherence in
interacting systems is often a result of their
synchronization~\cite{Pikovsky2001}.  It became a popular topic with
the discovery of synchronization of non-identical chaotic oscillators
\cite{Pecora1990}.  Over the years different types of synchrony were
studied, notably phase synchronization \cite{Rosenblum1996}.  There
were also numerous attempts to study more complicated interactions
under the names of generalized synchronization or interdependence
\cite{Rulkov1995, Pecora1995, Kocarev1996, Schiff1996, wojcik01da}.
In biological context synchronization is 
expected 
to play a major
role in cognitive processes in the brain \cite{Rosenblum2001,
  Singer1995, Varela2001} such as visual binding \cite{Singer1995} and
large-scale integration \cite{Varela2001}.  Various synchronization
measures were successfully applied to electrophysiological signals
\cite{Arnhold1999, Kaminski2001, Varela2001, Quiroga2002, Niebur2002a,
  Angelini2004, Lai2007, Korzeniewska2007}. In this work we
concentrate on nonlinear interdependence \cite{Arnhold1999,
  Quiroga2002}.

For an experimentalist it is often interesting to know how two systems
synchronize during short periods of evoked activity \cite{Fell2001,
  Kramer2004a}. Such questions arise naturally in analysing data from
animal experiments \cite{Kublik2003, Kublik2004, Swiejkowski2007,
  Wrobel2007}. One measures there electrical activity on different
levels of sensory information processing and aims at relating changes
in synchrony to the behavioral contex, such as attention or arousal.
It may be the case that the stationary dynamics (with no sensory
stimulation) corresponds to a fixed point.  For instance, when one
measures the activity in the barrel cortex of a restrained and
habituated rat, the recorded signals seem to be
noise~\cite{Kublik2003, Kublik2004, Swiejkowski2007}.  On the other
hand transient activity evoked by specific stimuli seems to provide
useful information. For example, bending a bunch of whiskers triggers
non-trivial patterns of activity (evoked potentials, EPs) in both the
somatosensory thalamic nuclei and the barrel
cortex~\cite{Swiejkowski2007, Leski2007}.

Explorations described in this paper aim at solving the following
problem.  Suppose we have two pairs of transient signals, for example
recordings of evoked potentials from thalamus and cerebral cortex in
two behavioral situations~\cite{Kublik2003, Swiejkowski2007}.  Can we
tell in which of the two situations the strength of coupling between
the structures is higher?  Thus we investigate if one can measure
differences in the strength of coupling between two structures using
nonlinear interdependence measures on an ensemble of EPs.  Since EPs
are short, transient signals, straightforward application of the
measures motivated by studies of systems moving on the attractors
(stationary dynamics) is rather doubtful and a more sophisticated
treatment is needed~\cite{Andrzejak2006, Kramer2004a}. Our approach is
similar in spirit to that advocated by Janosi and Tel for the
reconstruction of chaotic saddles from transient time
series~\cite{Janosi1994a}. (Note that the transients we study should
not be confused with the transient chaos studied by Janosi and Tel.)
Thus we cut pieces of the recordings corresponding to well-localized
EPs and paste them together one after 
another. Since we are interested
in the coupled systems, unlike Janosi and Tel, we obtain two
artificial time-series to which we then apply nonlinear
interdependence measures and linear correlations. It turns out that
this approach allows to extract the information about the strength of
the coupling between the two systems.

We test our method on a population model of information processing in
\begin{figure}[htbp]
  \includegraphics[width=0.35\textwidth]{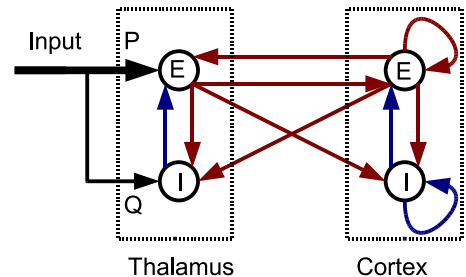}
  \caption{\label{schfig}Structure of  the model of the thalamocortical 
    loop used in the simulations.}
\end{figure}
thalamocortical loop (Figure~\ref{schfig}) consisting of two coupled
Wilson-Cowan structures~\cite{Wilson1972, Wilson1973}. Sensory
information is relayed through thalamic nuclei to cortical fields,
which in return send feedback connections to the thalamus. This basic
framework of the early stages of sensory systems is to a large extent
universal across different species and modalities
\cite{Shepherd2004}. 
To check that the results are  not specific to this particular  system we also study evoked dynamics of two coupled Rössler-type oscillators in non-chaotic regime.

The paper is organized as follows. In Sec.~\ref{sec:measures} we
define the measures to be used. In Sec.~\ref{sec:models} we describe
the models used to test our method.  Our model of thalamocortical loop
is discussed in Sec.~\ref{sec:models1} and a system of two coupled
Rössler-type oscillators is described in Sec.~\ref{sec:models2}. In
Sec.~\ref{sec:results} we present the results. In
Sec.~\ref{sec:results1} we show how various interdependence measures
calculated on the transients are related to the coupling between the
systems, while in Sec.~\ref{sec:results2} we study how the resolution
of our methods degrades with noise.  Finally, in
Sec.~\ref{sec:results3}, we apply time-resolved interdependence
measure $H_i$ \cite{Andrzejak2006} and compare its utility with our
approach. We summarize our observations in Sec.~\ref{sec:concl}.

\section{\label{sec:measures}Synchronization measures}

In the present paper we mainly study the applicability of nonlinear
interdependence measures on the transients. These measures, proposed
in~\cite{Arnhold1999}, are non-symmetric and therefore can provide
information about the direction of driving, even if the interpretation
in terms of causal relations is not
straightforward~\cite{Quiroga2000}.

These measures are constructed as follows. We start with two time
series $x_n$ and $y_n$, $n=1,\ldots, N$, measured in systems $\mrX$
and $\mrY$. We then construct $m$-dimensional delay-vector embeddings
\cite{Takens1980} $\bbx_n=(x_n,\ldots, x_{n-(m-1)\tau})$, similarly
for $\bby_n$, where $\tau$ is the time lag.  The information about the
synchrony is inferred from comparing the size of a neighborhood of a
point in $m$-dimensional space in one subsystem to the spread of its
equal-time counterpart in the other subsystem. The idea behind it is
that if the systems are highly interdependent then the partners of
close neighbors in one system should be close in the other system.
Several different measures exploring this idea can be considered
depending on how one measures the size of the neighborhood. These
variants include measures denoted by $S$, $H$ \cite{Arnhold1999}, $N$
\cite{Quiroga2002}, $M$ \cite{Andrzejak2003}. We have studied the
properties of most of these measures but for the sake of clarity here
we report only the results for the ``robust'' variant $H$ and a
normalized measure $N$, as they proved most useful for our purposes.

Let us, following~\cite{Arnhold1999}, for each $\bbx_n$ define a
measure of the spread of its neighborhood equal to the mean squared
Euclidean distance:
\[
R^{(k)}_n(\mrX) = \frac{1}{k}\sum_{j=1}^k (\bbx_n - \bbx_{r_{n,j}})^2,
\]
where $r_{n,j}$ are the time indices of the $k$ nearest neighbors of
$\bbx_n$, analogously, $s_{n,j}$ denotes the time indices of the $k$
nearest neighbors of $\bby_n$. To avoid problems related to temporal
correlations \cite{Theiler1986a}, points closer in time to the current
point $\bbx_n$ than a certain threshold are typically excluded from
the nearest-neighbor search (Theiler correction). Then we define the
$\bby$-conditioned mean
\[
R^{(k)}_n(\mrX|\mrY) = \frac{1}{k}\sum_{j=1}^k (\bbx_n -
\bbx_{s_{n,j}})^2,
\]
where the 
indices $r_{n,j}$ of the nearest neighbors of $\bbx_n$ are replaced
with the 
indices $s_{n,j}$ of the nearest neighbors of $\bby_n$. The
definitions of $R^{(k)}_n(\mrY)$ and $R^{(k)}_n(\mrY|\mrX)$ are
analogous.  The measures $H$ and $N$ use the mean squared distance to
random points:
\[
R_n(\mrX) = \frac{1}{N-1} \sum_{j\neq n} (\bbx_n - \bbx_j)^2,
\]
and are defined as
\begin{eqnarray*}
  H^{(k)}(\mrX|\mrY) & = & 
  \frac1N \sum_{n=1}^N \log \frac{R_n(\mrX)}{R^{(k)}_n(\mrX|\mrY)},\\
  N^{(k)}(\mrX|\mrY) &=& \frac1N \sum_{n=1}^N 
  \frac{R_n(\mrX)-R^{(k)}_n(\mrX|\mrY)}{R_n(\mrX)}.
\end{eqnarray*}
The interdependencies in the other direction $H^{(k)}(\mrY|\mrX)$,
$N^{(k)}(\mrY|\mrX)$ are defined analogously and need not be equal
$H^{(k)}(\mrX|\mrY)$, $N^{(k)}(\mrX|\mrY)$.

\begin{figure}[htbp]
  \includegraphics[width=0.47\textwidth]{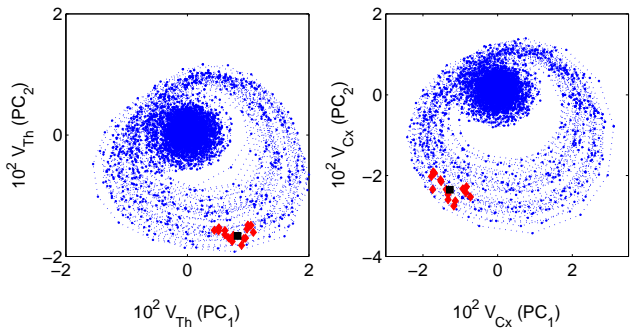}  
  \caption{\label{nbsfig}Delay-vector embeddings (shown in planes
    defined by the first two principal components) of pseudo-periodic
    signals obtained by gluing 50 evoked potentials generated in a
    model of thalamocortical loop. On the left (signal from
    ``thalamus'') a point is chosen (black square) and its 15 nearest
    neighbors are marked with red (gray) diamonds. On the right (``cortex'')
    the equal-time partners of the marked points from the left picture
    are shown. }
\end{figure}

Such measures base on repetitiveness of the dynamics: one expects that
if the system moves on the attractor the observed trajectory visits
neigborhoods of every point many times given sufficiently long
recording. The same holds for the reconstructed dynamics. However, if
the stationary part of the signal is short or missing, especially if
we observe a transient such as evoked potential, this is not the
case. Still, if we have noisy dynamics, every repetition of the
experiment leads to a slightly different probing of the neighborhood
of the noise-free trajectory. This observation led us to an idea of
gluing a number of repetitions of the same evoked activity (with
different noise realizations) together and using such pseudo-periodic
signals as we would use trajectories on a chaotic attractor. A similar
idea was used by Janosi and Tel in a different context for a different
purpose~\cite{Janosi1994a}. An example of a delay embedding of a
signal obtained this way is presented in Fig.~\ref{nbsfig}. Note that
artifacts may emerge at the gluing points. This is discussed
in~\cite{Janosi1994a}, and some countermeasures are proposed. For
simplicity we proceed with just gluing as we expect that the artifacts
only increase the effective noise level. The influence of noise is
studied in Sec.~\ref{sec:results2}.

Recently, time-resolved variants of the methods described above were
studied~\cite{Kramer2004a, Andrzejak2006}. They are applied to
ensembles of simultaneous recordings, each consisting of many
different realizations of the same (presumably short) process.  Let us
denote the $n$-th state vector in $j$-th realization of the
time-series by $\bbx^j_n$ ($\bby^j_n$, respectively), $j=1,\ldots, J$.
The idea in \cite{Kramer2004a} is, for given $\bbx^j_n$ to find one
neighbor in each of the ensembles. Then a measure (denoted $T$) based
on distances to these neighbors is constructed.  The proposition of
\cite{Andrzejak2006} is to look not at the nearest neighbors of a
given $\bbx_n$ no matter what time they occur at, but rather at the
spread of state-vectors at the same latency across the ensemble.  In
Sec.~\ref{sec:results3} we study the measure $H_i$ as defined in
\cite{Andrzejak2006}. Let $r_{i}^{j,l}$ denote the ensemble index of
the $l$-th nearest neighboor of $\bby_n^j$ among the whole ensemble
$\{\bby^j_n\}^{j=1,\ldots J}$. Define the quantities
\begin{eqnarray*}
  R^{j,(k)}_i(\mrX|\mrY) &=& \frac1k \sum_{l=1}^k (\bbx_i^j - 
  \bbx_i^{r^{j,l}_i})^2,\\
  R^{j,(k)}_i(\mrX) &=& \frac{1}{J-1} \sum_{s\neq j} (\bbx_i^j - 
  \bbx_i^s)^2.
\end{eqnarray*}
The time-resolved interdependence measure is further defined as
\[
H^{(k)}_i(\mrX|\mrY) = \frac1J \sum_{j=1}^J \log
\frac{R^{j,(k)}_i(\mrX)}{R^{j,(k)}_i(\mrX|\mrY)}.
\] 
Analogously one can define $H^{(k)}_i(\mrY|\mrX)$ and also
time-resolved variants of other interdependence measures.

In the numerical experiments described in this paper we use the
following parameters of the nonlinear interdependence measures: time
lag for construction of delay-vectors: $\tau = 1$, embedding dimension
$m=10$, number of nearest neighbors $k = 15$, Theiler correction $T=
5$. To calculate the interdependencies we used the code by Rodrigo
Quian Quiroga and Chee Seng Koh available at
\url{http://www.vis.caltech.edu/~rodri/Synchro/Synchro_home.htm}.  In
case of the measure $H_i$ we use the same embedding dimension and time
lag; here $k=1$.  To calculate this measure we used the code provided
in supplementary material to \cite{Andrzejak2006}.  To compare the
linear and nonlinear analysis methods we calculated the
cross-correlation coefficients using Matlab.

While in numerical studies the correctness of reconstruction can often
be easily checked by comparison with original dynamics, in analysis of
experimental data it can be a complex issue. Correct reconstruction is
a prerequisite for application of our technique.
For technical details on best practices of delay
embedding reconstructions, pitfalls and caveats, see~\cite{Kantz2004}.

\section{\label{sec:models}Model data}

\subsection{\label{sec:models1}Connected Wilson-Cowan aggregates}

Our model of the thalamocortical loop was based on the Wilson and
Cowan mean-field description of interacting populations of excitatory
and inhibitory neural cells~\cite{Wilson1972, Wilson1973}. In the
simplest version, which we used, each population is described by a
single variable standing for its mean level of activity
\begin{equation}
\begin{split}
  \tau_E \frac{\dd E}{\dd t} & =  
  -E + (k_E-r_E E)\mS_E (c_{EE} E - c_{IE} I +P), \\
  \tau_I \frac{\dd I}{\dd t} & = 
  -I + (k_I-r_I I)\mS_I (c_{EI} E - c_{II} I +Q).
\end{split}
\end{equation}
The variables $E$ and $I$ are the mean activities of excitatory and
inhibitory populations, respectively, and form the phase space of a
localized neuronal aggregate.  The symbols $\tau$, $k$, $r$, $c$
denote parameters of the model, $\mS$ are sigmoidal functions, $P$ and
$Q$ are input signals to excitatory and inhibitory populations,
respectively.
These equations take into account the absolute refractory period of
neurons which is a short period after activation in which a cell
cannot be activated again.  Such models exhibit a number of different
behaviors (stable points, hysteresis, limit cycles) depending on the
exact choice of parameters \cite{Wilson1972, Wilson1973}.  To relate
the simulation results to the experiment~\cite{Kublik2003,
  Swiejkowski2007} we considered the observable $V = E-I$, since the
electric potential measured in experiments is related to the
difference between excitatory and inhibitory postsynaptic potentials
(see the discussion in~\cite{Wilson1972}).

We studied a model composed of two such mutually connected aggregates,
which we call ``thalamus'' and ``cortex'' (Figure~\ref{schfig}).  Note
that the parameters characterizing the two parts are different (see
the Appendix~\ref{appA} for a complete specification of the model).
Specifically, there are no excitatory-excitatory nor
inhibitory-inhibitory connections in the thalamus.  Only the thalamus
receives sensory input, and we assume that $Q$ is always a constant
fraction of $P$.  The connections between two subsystems are
excitatory only.

To model the stimulus we assumed that the input ($P, Q$) switches at
some point from 0 to a constant value ($P_C, Q_C$), and after a short
time (on the time-scale of relaxation to the fixed point) switches
back to zero. This is clearly another simplification, as the real
input, which could be induced by bending a bunch of
whiskers~\cite{Kublik2003, Kublik2004, Swiejkowski2007}, would be a
more complex function of time.  However, the transient nature of the
stimulus is preserved.  In this simple setting we can understand that
the ``evoked potential'' corresponds to a trajectory approaching the
asymptotic solution of the ``excited'' system (with the non-zero input
$P_C, Q_C$), followed by a relaxation to the ``spontaneous activity''
in the system with null input.

\begin{figure}[htbp]
  \includegraphics[width=0.47\textwidth]{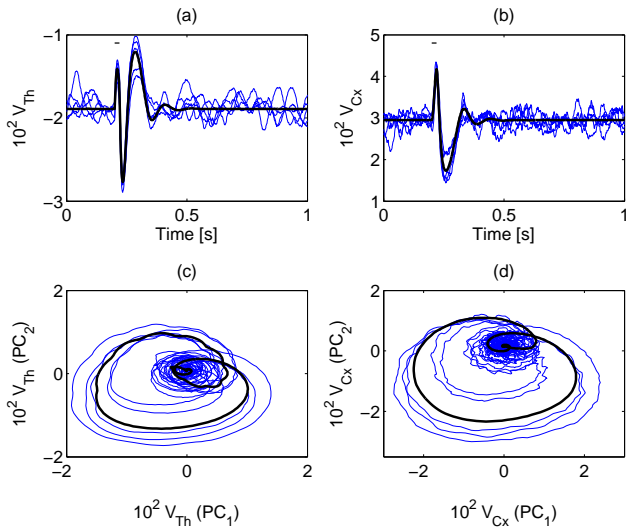}  
  \caption{ \label{epfig}``Evoked potentials'' ($V=E-I$), (a), (b) and
    their delay-vector embeddings  shown in a plane defined
	 by the
    first two principal components (c), (d). Plots (a) and (c): 
    thalamus, (b) and (d): cortex. The intervals above the EP 
	 indicate the duration of the non-zero stimulus. Black (thick) lines
    are solutions for the system without noise, blue (thin) curves are five
    different realisations of noisy dynamics.}
\end{figure}

The model parameters were chosen so that its response to brief
stimulation were damped oscillations of $V$ both in the thalamus and
the cortex similar to those observed in the experiments, both in terms
of shape and time duration~\cite{Kublik2003, Kublik2004,
  Swiejkowski2007} (Figure~\ref{epfig}).  However, apart from that, we
exercised little effort to match the response of the model to the
actual activity of somatosensory tract in the rat brain.  Our main
goal in the present work was establishing a method of inferring
coupling strength from transients and not a study of the rat
somatosensory system.  For this reason it was convenient to use a very
simplified, qualitative model. Interestingly, the response of the
model, measured for example as the activity of excitatory cells in the
thalamus, extends in time well beyond the end of the stimulation
(Figure~\ref{epfig}).  Such behavior is not observed in a single
aggregate and requires at least two interconnected
structures~\cite{Wilson1973}.

\begin{figure*}[htbp]
    \includegraphics[scale=0.9]{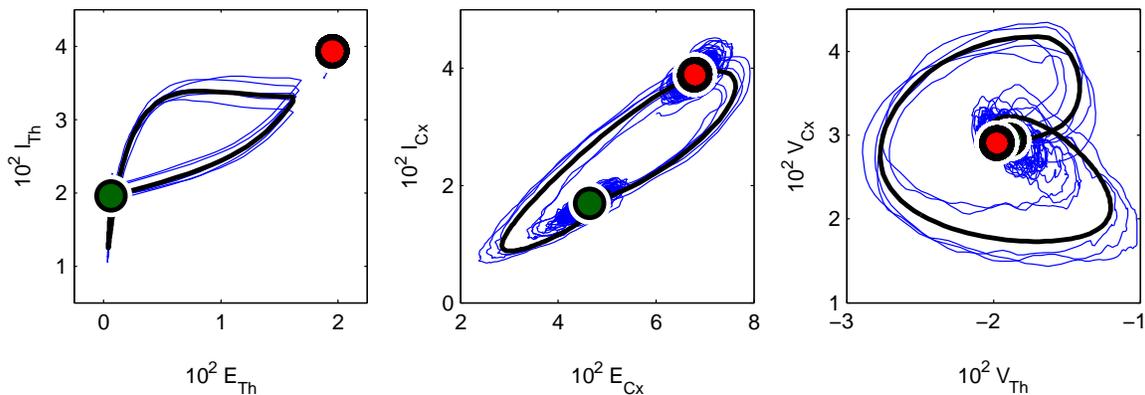}
    \caption{ \label{dynfig}Dynamics of the model. The green (lower
      left) and red (upper right) dots are fixed points in case of
      null or constant stimulation respectively, the black (thick) line is the
      noise-free transient dynamics. Blue (thin) lines are example
      trajectories of the model in the presence of noise. The plots
      show projections of the same dynamics to different planes.}
\end{figure*}

We performed numerical simulations in three modes: either stationary
(null or constant input), or not (transient input). The dynamics of
the model is presented in Figure \ref{dynfig}.
In case of transient input the simulation was done for $-1000\le
t\le1000$ms.  We used the stimulus $P$ and $Q$ which was 0 except for
the time $200<t<220$ when it was $P_C=3.5$ and $Q_C=0.3$. The system
settled in the stationary state during the initial segment ($t<195$)
which was discarded from the analysis.  The noise was simulated as
additional input to each of the four populations, see the
Appendix~\ref{appA} for the equations.  For each population we used
different Gaussian (mean $\mu=0$, standard deviation $\sigma =0.025$)
white noise, sampled at 1kHz and interpolated linearly to obtain
values for intermediate time points.  In case of stationary dynamics
we simulated longer periods, $-1000\le t\le20000$ms. The signals were
sampled at 100Hz before the synchronization measures were applied.

In case of constant or null stimulation the system approaches one of
the two fixed-point solutions which are marked by large dots in
Figure~\ref{dynfig}.  For the amount of noise used here the dynamics
of the system changes as expected: the fixed points become diffused
clouds (Figure \ref{dynfig}).  During the transient --- ``evoked
potential'' --- the switching input forces the system to leave the
null-input fixed point, approach the constant-input attractor, and
then relax back to its original state
(Figure~\ref{dynfig}). Of course, in the presence of noise the shape of the
transient is affected (Figure \ref{dynfig}).  Observe the similarity
between the embedding reconstructions of the evoked potentials (Figure
\ref{epfig}, bottom row) and the actual behavior in $V_\Th$-$V_\Cx$
coordinates (Figure~\ref{dynfig}, bottom row).

\subsection{Coupled Rössler-type oscillators}
\label{sec:models2}

\begin{figure}[htbp]
  \includegraphics[width=0.47\textwidth]{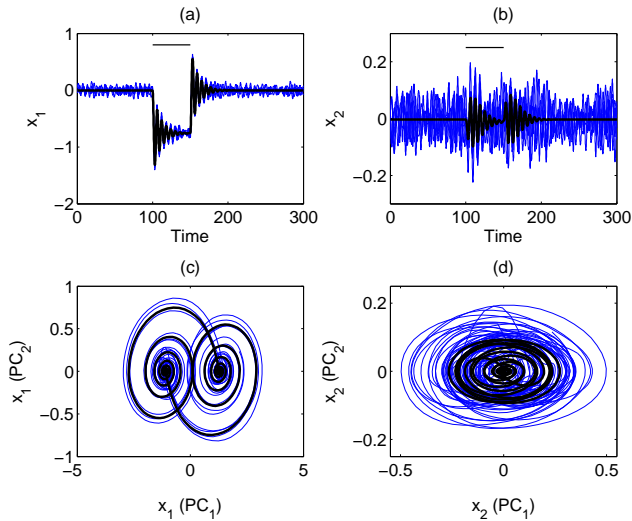}
  \caption{ \label{rosdynfig}(a), (b): signals ($x$ coordinate) from
    coupled Rössler-type oscillators; (c), (d): their delay-vector
    embeddings, shown in a plane defined by the first two principal
    components.  The intervals in (a) and (b) indicate the duration of
    non-zero input $P$. Black (thick) lines are solutions for the system
    without noise, blue (thin) curves are five different realisations of
    noisy dynamics.}
\end{figure}

While we are specifically interested in the dynamics of
thalamocortical loop which dictated our choice of the studied system,
we checked if our approach is not specific to this model. Our second
model of choice consisted of two coupled Rössler-type
oscillators~\cite{Rosenblum1996,Letellier2006} 
\begin{eqnarray*}
  \frac{\dd x_1}{\dd t} & = &-(1+\Delta\omega) y_1 - z_1 +\alpha
  C(x_2-x_1) + \xi_1,\\ 
  \frac{\dd y_1}{\dd t} & = &(1+\Delta\omega) x_1 -0.15y_1+ P + \xi_2,\\
  \frac{\dd z_1}{\dd t} & = &0.2 + z_1(x_1 -10) + \xi_3,\\
  \frac{\dd x_2}{\dd t} & = &-(1-\Delta\omega) y_2 - z_2 +\alpha
  C(x_1-x_2) + \xi_4,\\ 
  \frac{\dd y_2}{\dd t} & = &(1-\Delta\omega) x_2 -0.15y_2 + \xi_5,\\
  \frac{\dd z_2}{\dd t} & = &0.2 + z_2(x_2 -10) + \xi_6.
\end{eqnarray*}
We used the frequency detuning parameter $\Delta\omega = 0.05$ and the
maximum coupling constant $C=0.06$. The scaling parameter $\alpha$
took values from $0$ to $1$. The stimulation parameter $P$ was $0$
except for $200<t<250$ where it was set to $0.8$; the noise inputs
$\xi_i$, $i=1\ldots 6$ were Gaussian white noise with parameters as
for the Wilson-Cowan model. The simulation was done for $t\in[0,300]$
and segments from $t=195$ to $t=300$, sampled every $\Delta t =0.125$,
were used for the analysis of the transients. The synchronization was
measured between $x_1$ and $x_2$.  Parameters of the system were
chosen so that asymptotically it moved into a stable fixed point (note
the signs in the equations for ${y_1}$ and ${y_2}$) for both $P=0$ and
$P=0.8$. Therefore the transient dynamics (Fig.~\ref{rosdynfig})
is of the same type as in the model of thalamocortical loop: the
system switches briefly to the second stable point and then returns.
Note that the level of noise in the second subsystem is quite high and
the evoked activity is barely visible at the single trial level
(Fig.~\ref{rosdynfig}, right column).

\section{Results}
\label{sec:results}

\subsection{Inferring connection strength}
\label{sec:results1}

We aim at solving the following problem: suppose we have two pairs of
signals, for example recordings from thalamus and cerebral cortex in
two behavioral situations~\cite{Swiejkowski2007, Wrobel2007,
  Kublik2003, Kublik2004}.  Can we tell in which of the two situations
the strength of connections between the structures is higher? Thus we
need to find a measure being a monotonic function of the coupling
strength. We have studied this problem in our model of thalamocortical
loop (Section~\ref{sec:models1}). We scaled the strength of
connections from thalamus to cortex by changing $\alpha$ between 0 and
1, and calculated synchrony measures on signals from these
structures. The strength of connections from cortex to thalamus was
constant ($\beta = 1$); see the Appendix~\ref{appA} for the details.

\begin{figure*}[htbp]
  \begin{tabular}[t]{r}
    (a) \hfill \ \\ 
    \includegraphics[scale=0.9]{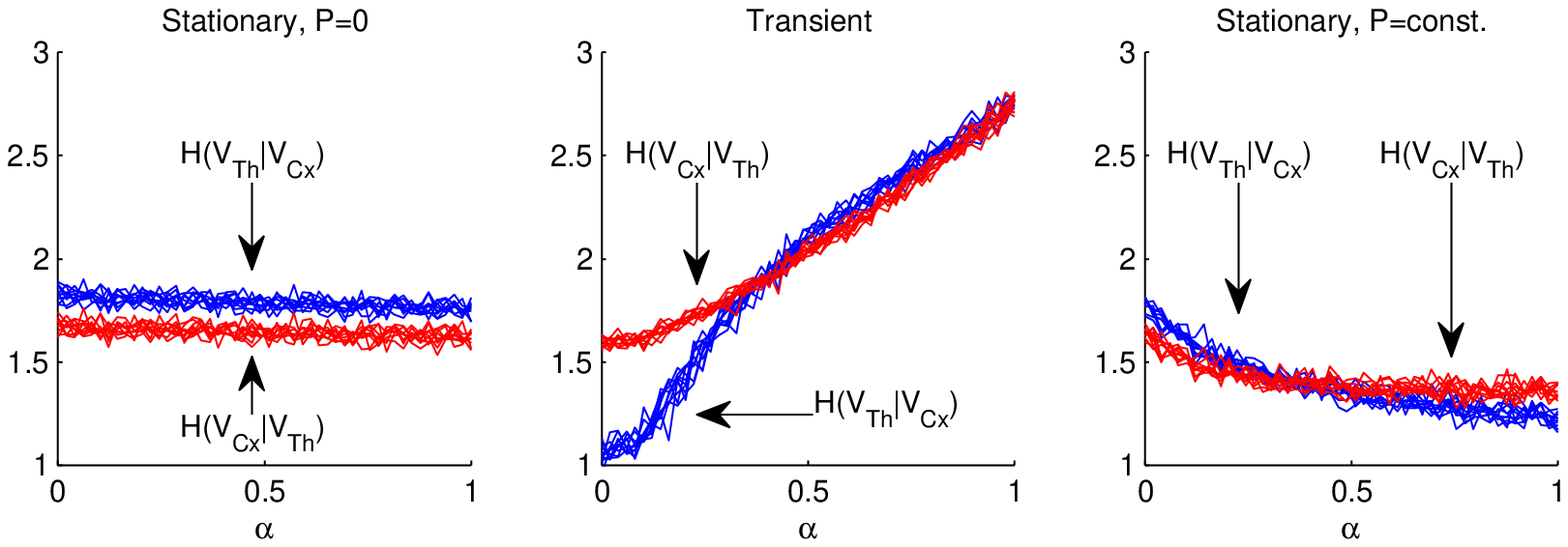}\\
    (b) \hfill \ \\ 
    \includegraphics[scale=0.9]{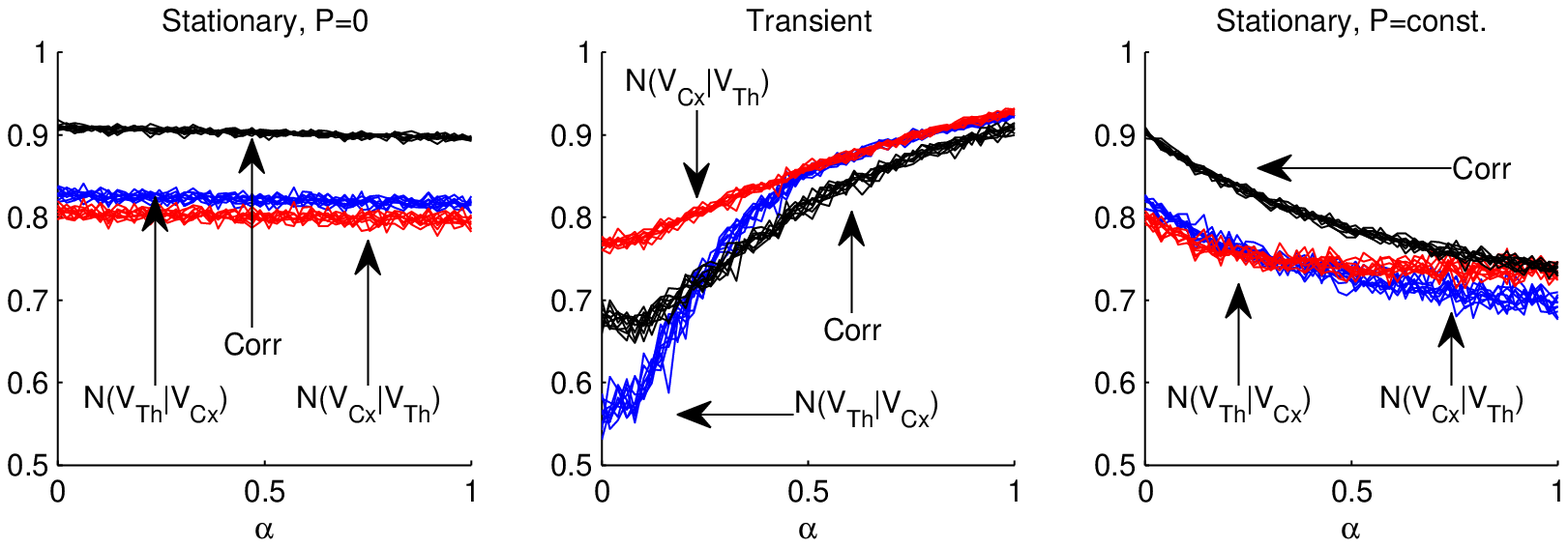}
  \end{tabular}
  \caption{ \label{synchrofig} Nonlinear interdependence measures (a)
    $H$, (b) $N$ and maximum of absolute value of cross-correlation
    coefficients for varying coupling constants.  Each measure is
    calculated for 10 independent realizations with different seeds.
    The parameter $\alpha$ scales the strength of connections from
    thalamus to cortex.}
\end{figure*}

Consider first stationary signals with $P=0$ or $P=\mathrm{const}$. Without noise the
system is in a fixed point and obviously it is impossible to obtain
the connection strength. However, given the noise, in principle the
dynamics in the neighborhood of the fixed point is also probed.  Thus
there is a possibility that the interdependence and the strength of
the coupling could be established during stationary parts of the
dynamics.  It turns out that for null stimulation neither the
interdependence measures nor the linear correlations detect any
changes in the coupling strength (Figure \ref{synchrofig}, left
column). For constant non-zero input there is a connection between the
coupling strength and the values of the measure but they are
anti-correlated and the dependence is not very pronounced (Figure
\ref{synchrofig}, right column). One must also bear in mind that while
it is possible to have no stimulation, in brain studies 
prolonged and constant stimulation in the present sense cannot be
experimentally realized (at least for most sensory systems) because of
the adaptation of receptors. The natural stimuli are necessarily
transient.

To use the synchrony measures on the transient we cut out pieces of
signal corresponding to the evoked potential, and pasted them one
after another. Thus obtained pseudo-periodic signal contained the same
underlying dynamics with each piece differing due to the noise. We
then applied the same measures as we did for the stationary signals.
In the simulations we calculated 50 ``evoked potentials''
(Figure~\ref{epfig}) for each value of $\alpha$. Plots in the middle
column of Figure~\ref{synchrofig} show the values of the
synchronization measures evaluated for different coupling strengths.
It can be seen that 
they are increasing functions of the coupling strength between the
subsystems. Therefore, our approach is indeed a viable solution to
the problem of data-based quantification of the coupling strength.

It is interesting to study the values of these interdependence
measures in different cases. Observe that
$H(V_{\Th}|V_{\Cx})>H(V_{\Cx}|V_{\Th})$ for $P=0$. The opposite is
true for transients (for small $\alpha$).  This is even more clearly
visible for $N$. 
In all the cases linear correlations showed similar trends to the
nonlinear measures $N(V_{\Th}|V_{\Cx}), N(V_{\Cx}|V_{\Th})$.

The asymmetry in the interdependence measures was originally intended
to be used for inferring the direction of the coupling or
driving. However, the inference of specific driving structure in every
case must follow a careful analysis of underlying dynamics (see, for
example, discussions in~\cite{Quiroga2000}
and~\cite{Arnhold1999}). Let us consider the plots in the middle
column of Figure~\ref{synchrofig}. For small $\alpha$ the dominant
connections are from the cortex to the thalamus so one might expect
that the state of the thalamus might be easier predictable from the
states of the cortex than the other way round. Thus one would intuitively expect
$H(V_{\Th}|V_{\Cx})>H(V_{\Cx}|V_{\Th})$. However, we observe the
opposite. The reason is that the measures used are related to the
relative number of degrees of freedom~\cite{Arnhold1999}. Loosely
speaking, as discussed~\cite{Quiroga2000}, the effective dimension of
the driven system (thalamus for small $\alpha$) is usually higher than
the dimension of the driver (which means that the response --- the
dynamics of the thalamus --- is ``more complex''). This effect is
further enhanced by the fact that we stimulate the thalamus in moments
unpredictable from the point of view of the cortex. Summarizing, the
result is compatible with the analysis in~\cite{Quiroga2000}. What
happens for higher $\alpha$ when the two measures become equal is
probably the coupling between the two subsystems becoming so strong
that the quality of prediction in any direction is comparable.

In the stationary case the situation is different as we observe the
asymptotic behavior. It turns out that for $P=0$ for every $\alpha$,
and for $P=\mathrm{const}>0$ for small $\alpha$ we have
$H(V_{\Th}|V_{\Cx})>H(V_{\Cx}|V_{\Th})$. But it seems that another
effect also plays a role here. The noise in the cortex has a higher
amplitude than in the thalamus and as a consequence it is easier to
predict the state of the thalamus from that of the cortex than in the
other direction. The reason for this disparity in the amplitudes is
the difference in the shape of the sigmoidal functions
$\mathcal{S}_q$. To summarize, here, the asymmetry of the measures
reflects internal properties of the two subsystems and not the
symmetry properties of the coupling between them.

\begin{figure*}[htbp]
\includegraphics[width=0.72\textwidth]{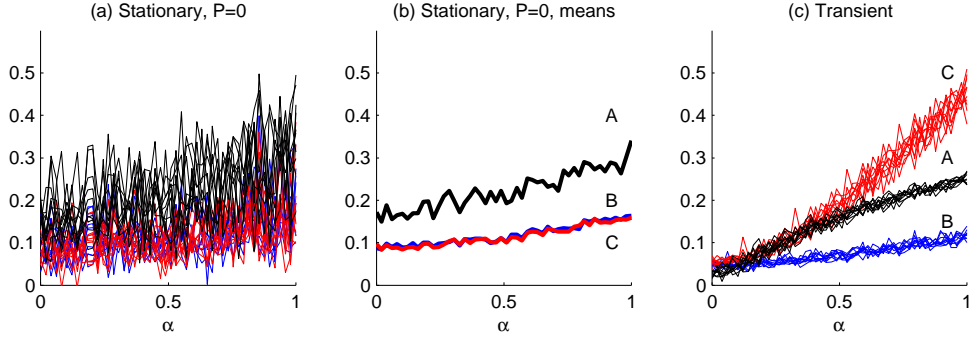}
\caption{ \label{rosfig}Nonlinear interdependence measures
  $H(\mrX_1|\mrX_2)$, $H(\mrX_2|\mrX_1)$ and maximum of absolute value
  of cross-correlation coefficients between signals from two
  symmetrically coupled Rössler-type systems with noise.  Coupling
  strength is proportional to $\alpha$. The panels (a) and (c) present
  results of 10 simulations with different seeds.  Additionally in
  stationary situation the means across repetitions are plotted for
  clarity in (b), the top curve (A) is cross-correlation. In
  stationary situation both nonlinear measures (B), (C) take similar values. On
  transients (c) $H(\mrX_2|\mrX_1)$ (C) is higher
  than $H(\mrX_1|\mrX_2)$ (B). The intermediate curves (A) are
  cross-correlation. }
\end{figure*}

Figure~\ref{rosfig} shows similar results obtained for two coupled
Rössler-type systems. In stationary situation the interdependence measures
are very noisy. Although a weak trend is visible, one would not be
able to reliably discriminate between, say, $\alpha = 0.25$ and
$\alpha=0.75$. The equality of the measures in two directions is due
to the fact that the systems are almost identical and symmetrically
coupled. 

If the interdependence is quantified on transient parts of the
dynamics, the situation improves considerably.  $H(\mrX_2|\mrX_1)$ has
a high slope and is a very good measure of the coupling strength
between the systems. Although $H(\mrX_1|\mrX_2)$ has a slope
comparable to that in the stationary case for $P=0$, the variability
of the results is much smaller, compared to the size of the
fluctuation in the ensemble mean in the stationary case.  The
difference between $H(\mrX_2|\mrX_1)$ and $H(\mrX_1|\mrX_2)$ reflects
the asymmetry of the driving (which makes the dynamics of $\mrX_1$
``more complex'' than the dynamics of $\mrX_2$), not of the coupling
(which is symmetric).

\subsection{Influence of noise}
\label{sec:results2}

The performance of the procedure described above depends on the level
of noise present in the system. To study this dependence we performed
the simulations of the thalamocortical model (the case of transient
dynamics) for 25\%, 50\%, 100\% and 200\% of the original noise
level. We found that for increasing level of noise the dynamics of the
system may change qualitatively: if the noise level is large enough
the system may be kicked out of the basin of attraction of the fixed
point and would not return there
after $P$ is reset to $0$. Instead it may fall into the basin of
attraction of another stable orbit or switch between the basins repeatedly. We observed
such behavior only once for 2500 simulations performed with 200\% of
the original noise and this trial was excluded from the analysis. Such
behavior becomes more frequent with increasing noise (e.g. 400\%) and
so we did not study this situation as it was very different from the
original dynamics of the system.

\begin{figure}[htbp]
\includegraphics[width=0.47\textwidth]{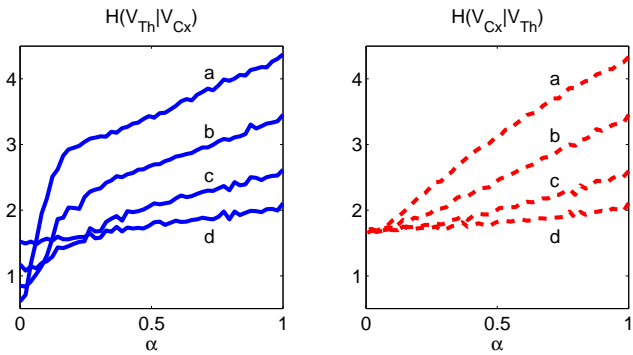}
  \caption{ \label{noisefig}Nonlinear interdependence $H$ in
    thalamocortical loop model for varying level of noise. The curves
    represent the values of the measure on transients. Noise level are
    25\% (a), 50\% (b), 100\% (c) and 200\% (d) of the original
    noise.}
\end{figure}

As one would expect, the higher the noise, the less sensitive the
measures are (Fig.~\ref{noisefig}).  However, even for twice the
original level of noise a weak trend in the interdependence is clearly
visible.

\subsection{Time-resolved measure $H_i$}
\label{sec:results3}

Since we are interested in the dynamics of non-autonomous systems one
might wonder if time-resolved measures, such as $H_i$ introduced
in~\cite{Andrzejak2006}, would not perform better in the problem of
inferring connection strength. We performed tests on cut-and-pasted
transient signals. This problem is different from the one studied
in~\cite{Andrzejak2006}.
\begin{figure}[htbp]
  (a) \hfill\strut \\
  \includegraphics[width=0.48\textwidth]{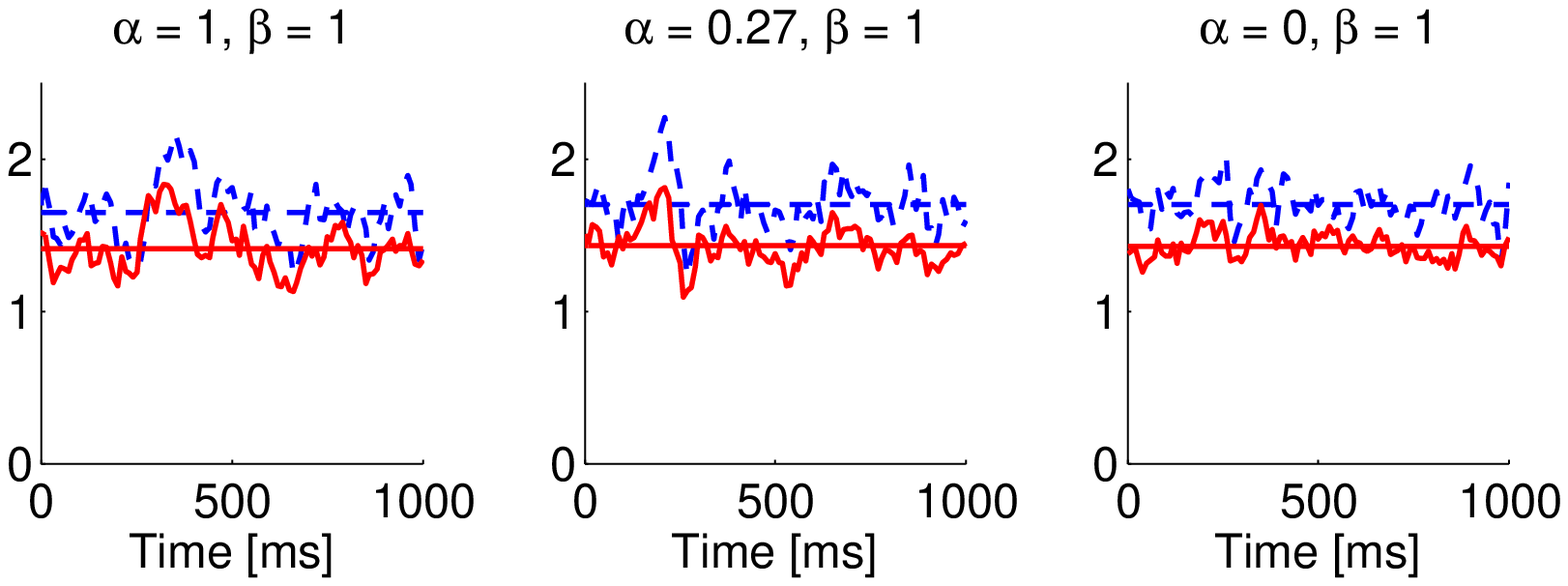}\\
  (b) \hfill\strut \\
  \includegraphics[width=0.48\textwidth]{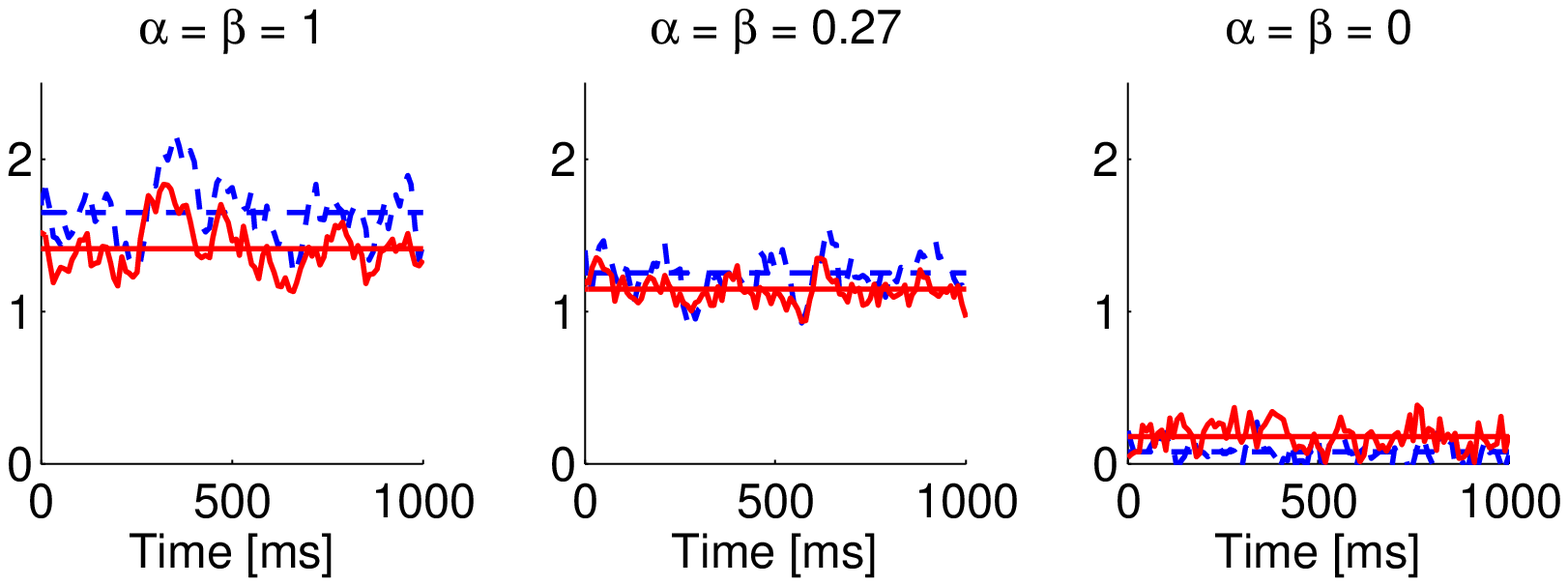}
  \caption{ \label{Hi1fig}Time-resolved nonlinear interdependence
    measure $H_i(V_{\Th}|V_{\Cx})$ (dashed lines) and
    $H_i(V_{\Cx}|V_{\Th})$ (solid lines) for 3 different values of the
    coupling $\alpha$.  The means are shown with horizontal lines. 
    (a) $\beta=1$,
    (b) $\beta=\alpha$.}
\end{figure}
There, two Lorenz systems were coupled for short periods of time and
$H_i$ was shown to identify these times of coupling well. In our
problem the coupling is constant in time, it is only the input to the
system that is varying. For the problem at hand the values of $H_i$ do
not seem to change with varying coupling constant $\alpha$
(Fig.~\ref{Hi1fig}, (a)) when $\beta$ is constant, $\beta=1$.  The
reason for this may be that even for $\alpha=0$ the subsystems are
coupled through the connections from cortex to thalamus.  This
hypothesis can be tested in another experiment, where all the
connections between the subsystems are scaled and $\alpha=\beta$.
\begin{figure}[htbp]
  \includegraphics[width=0.47\textwidth]{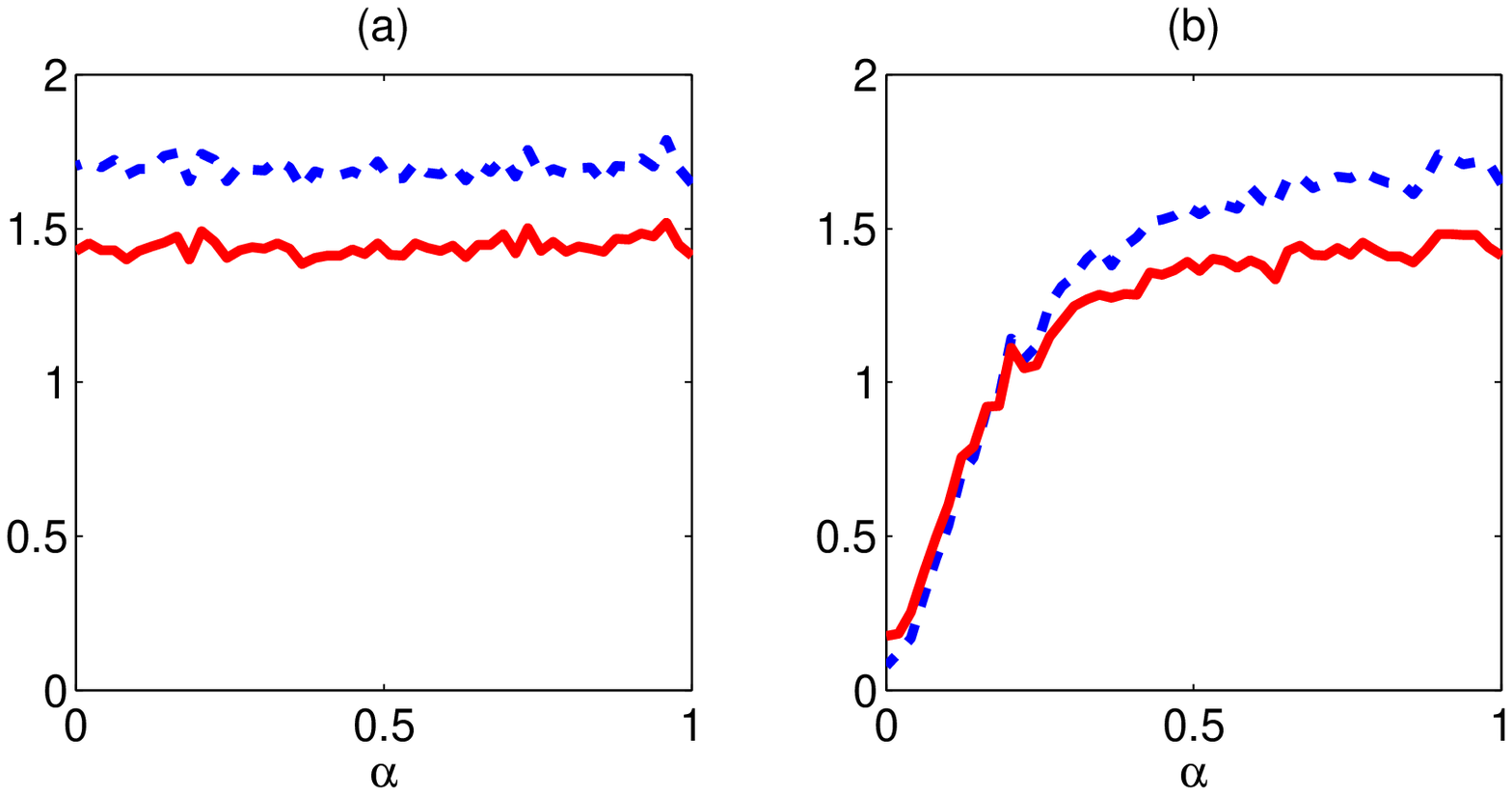}
  \caption{ \label{Hi6fig}Mean value of the time-resolved
    interdependence measure $H_i(V_{\Th}|V_{\Cx})$ (dashed lines) and
    $H_i(V_{\Cx}|V_{\Th})$ (solid lines). (a) $\beta=1$,
    (b) $\beta=\alpha$}
\end{figure}
Indeed, in this setup the measure $H_i$ is sensitive to the coupling
strength (Fig.~\ref{Hi1fig}, (b); Fig.~\ref{Hi6fig}).

One may also note that $H_i(V_{\Th}|V_{\Cx})$ is on average higher
than $H_i(V_{\Cx}|V_{\Th})$, exactly as for $H$ in case of $P=0$ and
con\-trary to what is observed using $H$ on transients
(Fig.~\ref{synchrofig} (a)). Thus it seems that for the problem of
inferring coupling strength between two systems the optimal approach
is to use $H$ or $N$, or linear correlations, on the transients, as
described in Section~\ref{sec:results1}.

\section{Conclusions}
\label{sec:concl}

To summarize, we have proposed a general approach for inference of the
coupling strength using transient parts of dynamics.  We have shown
that our approach gives more information about the coupling between
subsystems than the approach using the stationary part of dynamics in
case when the asymptotic dynamics is on a fixed point.  We have
checked the validity of this approach on a model of a thalamocortical
loop of sensory systems and on two coupled Rössler-type oscillators.
We showed that our method is quite robust with respect to increasing
level of noise as long as the dynamics does not change  qualitatively.
We have also shown that this method measures different aspects of
coupling than a time-resolved measure $H_i$ and than linear
correlations. We believe that our approach will be of use in many
other physical systems studied in the stimulus-response paradigm,
especially in the experimental context.

The results of Section~\ref{sec:results1} are compatible with our
preliminary studies of data from real neurophysiological
experiments~\cite{Swiejkowski2007}. There one cannot discern coupling
strength in two contextual situations basing on stationary recordings,
but the analysis of transients leads to clear differences between two
variants of experiment. The results of this analysis will be published
elsewhere.

\begin{acknowledgments}
  We are grateful to Ewa Kublik, Daniel Świejkow\-ski and Andrzej
  Wróbel for discussion of these topics. Some phase-space embeddings
  were calculated with ``Chaotic Systems Toolbox'' procedure {\tt
    phasespace.m} by Alexandros Leontitsis. This research has been
  supported by the Polish Ministry of Science and Higher Education
  under grants N401 146 31/3239, PBZ/MNiSW/07/2006/11 and
  COST/127/2007.  SŁ was supported by the Foundation for Polish
  Science.
\end{acknowledgments}

\appendix*
\section{\label{appA}Parameters of the models}
We use the following equations for the model of thalamocortical loop:
\begin{eqnarray*}
    \tau \frac{\dd E_{\Th}}{\dd t} &= & -E_{\Th}
    + (k_{E_{\Th}}-r {E_{\Th}}) \\* &&\times
    \mS_{E_{\Th}} (P - c_{1} I_{\Th} +\beta e_1 E_{\Cx} + \xi_1), 
\end{eqnarray*}
\begin{eqnarray*}
    \tau \frac{\dd I_{\Th}}{\dd t} &= &
    -I_{Th} + (k_{I_{\Th}}-r {I_{\Th}})\\* &&\times
    \mS_{I_{Th}} (Q + c_{2} {E_{\Th}+\beta e_2 E_{\Cx}} + \xi_2
    ),
\end{eqnarray*}
\begin{eqnarray*}
    \tau \frac{\dd E_{\Cx}}{\dd t} & = & -E_{Cx}
    + (k_{E_{\Cx}}-r E_{\Cx})\\* &&\times
    \mS_{E_{\Cx}} (c_{3} E_{\Cx}  - c_{4} I_{\Cx}+\alpha e_3 E_{\Th} + \xi_3), 
\end{eqnarray*}
\begin{eqnarray*}
    \tau \frac{\dd I_{\Cx}}{\dd t} &= &
    -I_{\Cx} + (k_{I_{\Cx}}-r {I_{\Cx}})\\* &&\times
    \mS_{I_{\Cx}} (c_{5} E_{\Cx} - c_{6} {I_{\Cx}}+\alpha e_4 E_{\Th} + \xi_4),
\end{eqnarray*}
where
\[
\mS_q(x) = \frac{1}{1+e^{-a_q(x-\theta_q)}} - \frac{1}{1+e^{a_q
    \theta_q}},
\]
$q$ standing for $E_\Th, I_\Th, E_\Cx, I_\Cx,$ and $\xi_i$,
$i=1\ldots4$ are noise inputs. The normalizing constants $k_q$ are
defined as $k_q = 1-\frac{1}{1+e^{a_q \theta_q}}$.

In the numerical experiments we used the following parameter values:
\begin{align*}
c_1&=1.35 & c_2&=5.35 & c_3&=15\\
c_4&=15 & c_5&=15   & c_6&= 3\\
  e_1 &= 10 & e_2 &= 20 & e_3 &= 10\\
e_4 &= 5  &\tau &= 10 \mathrm{ms} & r   &= 1\\
a_{E_{\Th}} &= 0.55 &\theta_{E_{\Th}} &= 11 &a_{I_{\Th}} &= 0.25\\
 \theta_{I_{\Th}} &= 9
&a_{E_{\Cx}} &= 1
&\theta_{E_{\Cx}} &= 2\\
a_{I_{\Cx}} &= 2
&\theta_{I_{\Cx}} &= 2.5
\end{align*}

The strength of connections was scaled by $\alpha\in[0,1]$. Everywhere
except in Section \ref{sec:results3} we used $\beta=1$. In Section
\ref{sec:results3} we used either $\alpha\in[0,1]$ and $\beta=1$, or
$\alpha\in[0,1]$ and $\beta=\alpha$.

\bibliographystyle{apsrev}

\end{document}